\documentstyle[epsfig]{mn}

\title[Faraday Rotation Template]{A Faraday Rotation Template for the Galactic Sky}

\author[P. Dineen \& P. Coles]{Patrick Dineen\thanks{E-mail: ppxptd@nottingham.ac.uk} \& Peter Coles\\
School of Physics \& Astronomy, University of Nottingham, University Park, Nottingham, NG7 2RD, United Kingdom\\}

\begin{document}
\maketitle

\begin{abstract}
Using a set of compilations of measurements for extragalactic radio sources we construct
all-sky maps of the Faraday Rotation produced by the Galactic magnetic field. In order
to generate the maps we treat the radio source positions as a kind of "mask" and
construct combinations of spherical harmonic modes that are orthogonal on the masked
sky. As long as relatively small multipoles are used the resulting maps are quite stable
to changes in selection criteria for the sources, and show clearly the structure of the local 
Galactic magnetic field. We also suggest the
use of these maps as templates for CMB foreground analysis, illustrating the idea with
a cross-correlation analysis between the Wilkinson Microwave Anisotropy Probe (WMAP)
data and our maps. We find a significant cross-correlation, indicating the presence
of significant residual contamination.
\end{abstract}


\section{Introduction}

The origin of large-scale magnetic fields observed on galactic and cluster
scales is unknown. The magnetic fields, with observed strengths of $\sim\!10^{-6}$G, could be the consequence of an amplification of a tiny seed
($\la\!10^{-20}$G) by a dynamo mechanism. Alternatively, the compression of a primordial seed
($\sim\!10^{-9}$G) by protogalactic collapse could lead to the fields we see
today. Both scenarios require an initial primordial field. Furthermore, the
two mechanisms need to explain the high redshift magnetic fields observed in
galaxies \cite{kpz92} and damped Lyman-$\alpha$ clouds \cite{wlo92}. Magnetic
fields play a crucial role in star formation as well as possibly playing an
active role in galaxy formation as a whole \cite{w78,w02}. Thus, one of
the most significant tasks in cosmology is unravelling the mystery behind
magnetic fields; from the primordial field to our own Galactic field. 

The cosmic microwave background (CMB) provides us with the most distant and extensive probe of the early
universe. A primordial magnetic field will leave an imprint in this
radiation. Various methods have been developed that seek these
signatures. Barrow, Ferreira and Silk (1997) use the anisotropic expansion caused
by the presence of a homogeneous primordial field to place limits on its size
from large-angle CMB measurements. Others have sought correlation between
different scales in the temperature anisotropies \cite{cmkr04,ncov04} or computed the
effects the field has on the polarisation-temperature cross-correlation \cite{sf97,l04}.
The existence of a magnetic field at last-scattering also leads to a possible measurable Faraday
rotation of the polarised CMB light \cite{kl96}.

At the other end of the scale, investigation of our own Galaxy's magnetic field has led to the development of
a number of different techniques; Han (2004) provides a concise review of the subject. Techniques include observing Zeeman splitting, polarised
starlight, synchrotron radiation, Faraday rotated light and polarised dust emission. However, even
with all these methods, there are still outstanding problems. This has led to
a lack of consensus on key issues: from the number of spiral arms; how the
arms are
connected; to the direction the field takes along the arms \cite{v97,h04}.  

To fully understand magnetic fields, we need a coherent picture throughout
different epochs. Theories and models can then be tested against this observational picture.  A rotation measure (RM)
map of the full sky has the potential to fulfil such a goal. RM values
probe the integral of the magnetic field from the radiation 
source to the observer. Obviously, the information encoded in such a map depends on the location
of the radiation that is rotated. Faraday rotated polarised CMB radiation will offer both a picture of the primordial
field (at the surface of last scattering) and that of our Galaxy. The recent detection of CMB polarisation by DASI \cite{klpc02} and confirmation of this
via the Wilkinson Microwave Anisotropy Probe (WMAP; Kogut et al. 2003), have
opened up a new avenue in CMB research. 
Future results from WMAP and Planck satellites will offer polarised data covering the
full-sky over a range of frequencies (seven in the case of Planck).
By forming a RM map with the data and looking at differing scales, we should be
able to untangle the information in the data; on the largest scales
the local magnetic field can be studied, whereas the primordial
field can be studied on the smaller scales. 

On the other hand, one of the
main tools for probing local magnetic fields (such as our Galaxy's) involves
utilising RM values from extragalactic sources.
Catalogues containing RM values of extragalactic sources have been
used to map the Galactic field (eg. Frick et al. 2001). Combining this data
with rotation measures from pulsars that are located within our Galaxy, it is
conceivable that a 3-dimensional image of the Galactic magnetic field can be
built. However, current RM catalogues are both sparsely populated and unevenly
sampled. Thus, astute methods are required to produce a RM map with the data at hand.  

So, what do we intend to do? We attempt to map the RM values as a
function of angular position giving ${\cal R}(\Omega)$, where we use ${\cal
  R}$ to denote the
Faraday rotation measure and $\Omega$ for the angular position. Catalogues containing RM values of
extragalactic sources are used to construct the function. The observed
spatial distribution of the RM values can be expanded over a set of orthogonal basis functions. 
For analysis
of data distributed on the sky, expansion over spherical harmonics seems natural 
\begin{equation}
{\cal R}(\Omega)=\sum _{l=1}^{\infty }\sum _
{m=-l}^{m=+l}a_{l,m}Y_{l,m}(\Omega ),
\end{equation}
where the $a_{l,m}$ are the spherical harmonic coefficients and the $Y_{l,m}$ are the
spherical harmonics. The properties of the spherical harmonics are well
understood and the calculation of the $a_{l,m}$ will allow us to utilise
routines within the HEALPix\footnote{http://www.eso.org/science/healpix/} 
package \cite{healpix} for visualisation purposes 
and further analysis. However, a non-uniform distribution of data points
compromises spherical harmonic analysis due to the loss of
orthogonality \cite{g94}. It is more fruitful to analyse a system using orthogonal
functions: the statistical properties of the coefficients are simplified. If
nonorthogonal functions are used, the properties of the system and the basis
are confused. Therefore, we would like to construct an orthonormal basis with
functions closely related to those of the spherical harmonics. The spherical
harmonic coefficients can then be obtained from the resultant coefficients of
the orthonormal basis.  

Spherical harmonic analysis of extragalactic sources has been previously performed by Seymour
(1966,1984). However, the analysis was carried out using a different form of
orthogonalisation and only on a set of 65 sources. 

The RM map resulting from our method will be a useful tool for probing
Galactic magnetic structure. The map will also be a
valuable point of reference when investigating mechanisms that involve the Galactic magnetic field. For
example, CMB foregrounds (synchrotron, dust, free-free emission) are correlated with rotation measures \cite{dc04}. 
There is evidence of another foreground component \cite{dtdg04}, labelled foreground X, that is
spatially correlated with $100\mu$m dust emission. Spinning dust grains \cite{dl98}
are the most popular candidates for causing this anomalous emission. A RM map
can provide insight
into the role of these grains as they may align with the local magnetic
field. Foreground removal will be particularly challenging in CMB polarisation
studies as foregrounds are more dominant than in the temperature
anisotropies. Also, single
frequency polarisation measurements will not be able to remove the effects
of Faraday rotation through the Galactic magnetic field. Thus, the extent to which the results have been effected
by the $E$-mode signal rotating into the $B$-mode signal (and vice versa) is unknown. 
   
The layout of the paper is as follows. In the next section we describe the
three rotation measure catalogues used in our analysis. In describing the data
we clarify the meaning of extragalactic rotation measures. In Section \ref{sec:basis} we illustrate a method to
generate orthonormal basis functions for each catalogue.
From the coefficients of the new basis, the spherical
harmonic coefficients are calculated. In Section \ref{sec:results}
we present the resulting RM maps and discuss the observed features. In
Section \ref{sec:correlations} we give a brief application of the maps. Correlations
are sought between the RM maps and cleaned CMB-only maps. The conclusions are presented and
discussed in Section \ref{sec:conclusion}.

\section{Rotation measure catalogues}
\label{sec:data}

Faraday rotation measures of extragalactic radio sources are
direct tracers of the Galactic magnetic field. When
plane-polarised radiation propagates through a plasma with a
component of the magnetic field parallel to the direction of
propagation, the plane of polarisation rotates through an angle
\(\phi\) given by
\begin{equation}
\label{phi} \phi={\cal R}\lambda^{2},
\end{equation}
where the Faraday rotation measure is measured in $\mathrm{rad\,m}^{-2}$ where
 \begin{equation}
\label{rm} {\cal R}=\frac{e^{3}}{2\pi m^{2}_{e}c^4}\int
n_eB_\parallel \,ds.
\end{equation}
Note that $B_\parallel$ is the component of the magnetic field
along the line-of-sight direction. The observed RM of
extragalactic sources is a linear sum of three components: the
intrinsic RM of the source (often small); the value due to the
intergalactic medium (usually negligible); and the RM from the
interstellar medium of our Galaxy \cite{bmv88}. The latter
component is usually assumed to form the main contribution to the
integral. If this is true, studies of the distribution and
strength of RM values can be used to map the Galactic magnetic
field \cite{vk75}. Even if the intrinsic contribution were not
small, it could  be ignored if the magnetic fields in different
radio sources were uncorrelated and therefore simply add noise to
any measure of the Galactic field \cite{fsss01}. In a similar
vein, the distributions of RM values have been used to measure
local distortions of the magnetic field, such as loops and
filaments, and attempts have also been made to determine the
strength of intracluster magnetic fields \cite{ktk91}. In what
follows we shall use RM values obtained from three catalogues in an attempt to
map ${\cal R}(\Omega)$ over the whole sky. 

All three catalogues are sparsely populated and have non-uniform
distributions. It is essential to remove the structure due to the spatial
distribution of the sources. This structure is unique to each
catalogue. Therefore, for each catalogue a new set of orthonormal functions has to be generated. 

The three catalogues used are those of Simard-Normandin et al. (1981;
hereafter S81), Broten et al. (1988-updated in 1991; B88) and Frick et
al. (2001; F01). S81 present an all-sky catalogue of
rotation measures for 555 extragalactic radio sources (ie. galaxies and
quasars). B88 and F01 contain 674 and 800 sources respectively. In F01, the
two other catalogues are combined with smaller studies of specific regions in the
sky (see paper for details). They also provide slightly reduced versions of the other two
catalogues. Sources with significantly larger RM values than those in the other studies are removed,
leaving catalogues of 551 sources for S81 and 663 for B88. In our analysis we
will use these versions of the two catalogues.

Finally, we reject sources with ${\cal R} > 300\,\mathrm{rad\,m}^{-2}$. Such large 
RM values are unlikely to represent real features of the Galactic magnetic field: probably they arise from
incorrect determination of ${\cal R}$ due to the $n \pi$
ambiguity in polarisation angle; magnetic fields within the sources; Equation
(\ref{phi}) being incorrect; and so on. This final selection criteria reduces 
the catalogues to 540 sources for S81, 644 for B88, and 744 for F01.

\section{Generating a new basis}
\label{sec:basis}
We can only observe RM values where there happens to be a line of sight. This
means we see the RM sky through a peculiar ``mask''.
We wish to generate a new orthogonal basis that takes account of the spatial
structure of this mask. In particular, we need to find
a set of functions that are orthogonal on the incomplete sphere. Ideally, these new functions should be related to the spherical harmonics (which are
orthogonal functions on a complete sphere). This will enable us to determine
the spherical harmonic coefficients from the new functions and their
coefficients.

G\'orski (1994) tackles the problem from the point of view of CMB analysis. The
determination of the angular power spectrum is a crucial element of much work
in the field. If the temperature anisotropies form a Gaussian random field
then they can be completely characterised by the angular power spectrum. In order to obtain the
angular power spectrum, one needs to estimate the spherical harmonic
coefficients. At low Galactic latitudes ($b < \mathrm{20^o}$) foreground contamination
is severe. Therefore, it is preferable to obtain an estimate of the spherical
harmonic coefficients outside this region. G\'orski (1994) calculates a new set of
functions that are orthogonal to this cut sphere. These functions are used to
calculate the spherical harmonic coefficients and thus estimate the angular
power spectrum from the two-year COBE-DMR data \cite{bkhb94}.

In order to see how the method works it is prudent to look at the definition
of orthogonal functions. Let us consider two complex functions $A(x)$ and $B(x)$. If
\begin{equation}
\int_{a}^{b}A^{\ast}(x)B(x)\, dx = 0,
\end{equation}
then $A(x)$ and $B(x)$ are orthogonal over on the interval \{$a \, , \, b$\}. If we
incorporate these two functions into a vector {\bf v}=[$A(x)$,$B(x)$], the
orthogonality of the functions can be expressed through the scalar product
\begin{equation}
\langle {\bf v \cdot v}^T\rangle_{\{a,b\}}={\bf I}.
\end{equation}
We shall now look at orthogonal functions in the context of a complete sphere. A function can be
described by spherical harmonics $Y_{l,m}(\Omega)$ up to an order $l_{\mathrm{max}}$. We can form
an $(l_{\mathrm{max}}+1)^2 $--dimensional vector {\bf
  y}=[$Y_{0,0}(\Omega),Y_{1,-1}(\Omega),Y_{1,0}(\Omega),Y_{1,1}(\Omega),
\ldots,Y_{l_{\mathrm{max}},l_{\mathrm{max}}}(\Omega)$]. The scalar product is
then defined as 
\begin{equation}
\langle{\bf y \cdot y}^T\rangle_{\{\mathrm{full \: sky}\}}={\bf I}.
\end{equation}
The sky can therefore be fully described by 
\begin{equation}
{\cal R}(\Omega)=\sum_{i=1}^{(l_{\mathrm{max}}+1)^2} a_i Y_i(\Omega) \equiv {\bf a}^T \cdot {\bf y}.
\end{equation}
However, when the sphere is incomplete due to a Galaxy cut or a more
complex mask being applied, we have
\begin{equation}
\langle{\bf y \cdot y}^T\rangle_{\{\mathrm{cut \: sky}\}}={\bf W} \neq {\bf I},
\end{equation}
where {\bf W} is the coupling matrix. A new basis, where the equality is
true, can be constructed in the following manner. The procedure is a type of
Gram-Schmidt orthogonalisation. {\bf W} can be
Choleski-decomposed into a product of a lower triangular matrix {\bf L} and
its transpose
\begin{equation}
{\bf W}={\bf L \cdot L}^T.
\end{equation}
The inverse matrix ${\bf\Gamma}={\bf L}^{-1}$ is then computed. The new set of
functions on the cut sky is 
\begin{equation}
\label{psidef}
\psi=\langle {\bf \Gamma \cdot y} \rangle_{\{\mathrm{cut \: sky}\}}.
\end{equation}
By construction, we have
\begin{eqnarray}
\langle \psi \cdot \psi^T\rangle_{\{\mathrm{cut \: sky}\}} &=& {\bf \Gamma \cdot y \cdot
y}^T {\bf \cdot \Gamma}^T \nonumber\\ 
&=& {\bf \Gamma \cdot  L \cdot L}^T {\bf \cdot \Gamma}^T \nonumber\\
&=& {\bf L}^{-1} \cdot {\bf L \cdot L}^T \cdot ({\bf L}^{-1})^T \nonumber \\
&=& {\bf I}.
\end{eqnarray}
This can be a useful cross-check for testing the code. Finally, the new basis functions can be used to describe $\cal R$
\begin{equation}\label{eqn:series}
{\cal R}(\Omega)=\sum_{i=1}^{(l_{\mathrm{max}}+1)^2} c_i \Psi_i(\Omega) \equiv {\bf c}^T \cdot {\bf \psi}.
\end{equation}

At this point, only the angular positions $\Omega_n$ of the sources have been
required. In order to obtain the coefficients $c_i$ of the new basis, the
rotation measure ${\cal R}_n$ themselves are required. It is worth putting it
into the context of the data we
have. Let the number of sources in our catalogue be $N$ and let
$(l_{\mathrm{max}}+1)^2\!=\!M$. It should be evident that we have a set of
simultaneous equations which have to be solved in order to obtain the coefficients of the new
basis functions
\begin{eqnarray}\label{eqn:simultaneous}
{\cal R}(\Omega_1)&\!=\!&\! c_1\psi_1(\Omega_1)+c_2\psi_2(\Omega_1)+\ldots+c_M\psi_M(\Omega_1) \nonumber \\ 
{\cal R}(\Omega_2) &\!=\!&\! c_1\psi_1(\Omega_2)+c_2\psi_2(\Omega_2)+\ldots+c_M\psi_M(\Omega_2) \nonumber \\
\vdots &&\vdots  \nonumber \\
       {\cal R}(\Omega_N)&\!=\!&\! c_1\psi_1(\Omega_N)+c_2\psi_2(\Omega_N)+\ldots+c_M\psi_M(\Omega_N). 
\end{eqnarray}
So, as long as $N\!>\!M$, these equations should be solvable. Ultimately, we wish to
obtain the spherical harmonic coefficients $a_{l,m}$. Using Equation
(\ref{psidef}), we see
\begin{equation}
{\cal R} ={\bf a}^T \cdot {\bf y}={\bf c}^T \cdot {\bf \psi}= {\bf c}^T \cdot
{\bf \Gamma \cdot y} 
\end{equation}
and therefore
\begin{equation}
{\bf a}^T={\bf c}^T \cdot {\bf \Gamma} \;\rightarrow\; {\bf a} = {\bf \Gamma}^T
\cdot {\bf c}.
\end{equation}
So, we have obtained the spherical harmonic coefficients.

There are some practical points that have been glossed over in the above
description of the method. Firstly, the RM values in the catalogues need to be
smoothed. Otherwise, as the series in Equation (\ref{eqn:series}) is finite, we
will be attempting to fit large-scale waves to small-scale
features.
Ideally, the smoothing will take place in the new basis, however, this is
impractical. Therefore, we chose to smooth in harmonic space. Around each
source, a hoop of $\mathrm{20^o}$ is thrown and the average RM value is taken of the sources
within the hoop. This could have been done via a more sophisticated method,
say a Gaussian-weighted mean of RM values. However, we chose to use the simple
approach. The size of the hoop was chosen to match
$l_{\mathrm{max}}$ $(\sim 16)$ closely in angular size. Furthermore, it coincided with the
limiting resolution of the wavelet method used in Frick et al. (2001) on the
same catalogues. 

Secondly, we need to determine to what order we take the series up to,
i.e. the value of $l_{\mathrm{max}}$. We do this through trial and error. 
As we will expand upon in the next section, RM maps were generated for $l_{\mathrm{max}}$ values of 8, 10, 15, 16, 17 and
18. The power spectrum for each map was studied. 
At some limiting value of $l_{\mathrm{max}}$ the shape at low $l$ alters as
features become unstable. Maxima and minima points explode since we are trying to
fit more and more function to the same amount of data. That is to say, $N$ is
getting too similar to $M$. 

Finally, the convention for spherical harmonics has to be chosen carefully. The new
basis functions were calculated using the convention of G\'orski (1994) where
\begin{equation}\label{eqn:Gorskiharmonics}
Y_{l,m}(\theta,\phi)=\sqrt{ \left(\frac{2l+1}{2}\right)}\sqrt{\left(\frac{(l-|m|)!}{(l+|m|)!}\right)}P_l^{|m|}(\cos\theta)f(\phi)
\end{equation}
and $f(\phi)=\pi^{-1/2}\cos(m\phi),\;(2\pi)^{-1/2}$, or $\pi^{-1/2}\sin(|m|\phi)$
for $m>0,\;=0$, or $<0$. The spherical harmonic coefficients can then be trivially converted into those
that adhere to the HEALPix definition of the harmonics where now
$f(\phi)=(2\pi)^{-1/2}[\cos(m\phi)+i\sin(m\phi)]$ or
$(-1)^{m}(2\pi)^{-1/2}[\cos(m\phi)+i\sin(m\phi)]$ for $m \geq 0$ or $<
0$. The convention of G\'orski (1994) was chosen since the information within
the coefficients is more
highly compressed. Whereas in the HEALPix definition,
the coefficient are complex with a symmetry between +$m$ and -$m$, those following the convention of G\'orski (1994)
are real and contain no such symmetry. The redundant
information in the HEALPix coefficients leads to
confusion when solving Equation (\ref{eqn:simultaneous}).

\section{All-Sky RM maps and their Interpretation}
\label{sec:results}

\begin{figure} {
\centering{\epsfig{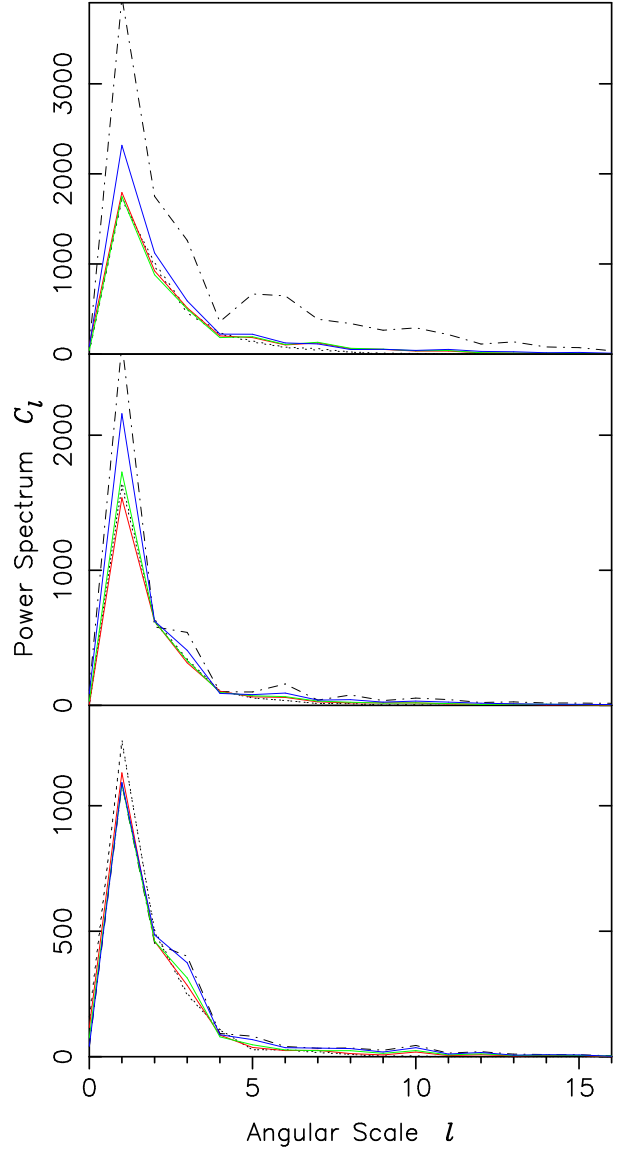}} 
}
 \caption{\label{fig:cl} Power spectra (measured in $\mathrm{rad}^{2}\,\mathrm{m}^{-4}$). Red $l_{\mathrm{max}}$=15, green $l_{\mathrm{max}}$=16, blue
 $l_{\mathrm{max}}$=17, dotted $l_{\mathrm{max}}$=8 and 10, and dash-dot $l_{\mathrm{max}}$=18. From {\it
 top} to {\it bottom}:
 S81, B88 and F01 catalogues.}
\end{figure}

For all three catalogues, sets of spherical harmonic coefficients were
calculated with $l_{\mathrm{max}}$ being set to 8, 10, 15, 16, 17 and 18. From these
coefficients, RM maps were produced using
the '{\tt synfast}' routine in the HEALPix package. To
see whether a RM map was
displaying real features or whether the series expansion had been extended too
far, we looked at the angular power spectrum of each map. The
angular power spectrum is the harmonic
space equivalent of the autocovariance function in real space. It is defined as 
\begin{equation}
C_l=\frac{1}{2l+1}\sum_m |a_{l,m}|^2.
\end{equation}
The spectra of the RM maps are shown in Figure \ref{fig:cl}. For all three
catalogues, it is clear
that the shape of the spectra is
consistent up to $l_{\mathrm{max}}=16$. Extending the series expansion to higher values
of $l_{\mathrm{max}}$ leads to fluctuations in the power on the largest-scales
(low $l$). The method finds it harder to reconcile the data with the
increasing number of basis functions. The result is
that maxima and minima explode as too much freedom is given. This is clearly
visible in the maps for $l_{\mathrm{max}}$=17 and 18 (not displayed). Although
looking at the bottom sets of spectra for F01, we see the spectra for $l_{\mathrm{max}}$=17 
is consistent until the octupole ($l$=3) where it spikes. 
This suggest that due to the larger data size of F01 it is more able to cope
with the demands of increasing the series expansion. Therefore, at times
throughout this section, we will focus our analysis on the RM map from the F01
catalogue with $l_{\mathrm{max}}$=16. 

\begin{figure} {
\centering{\epsfig{file=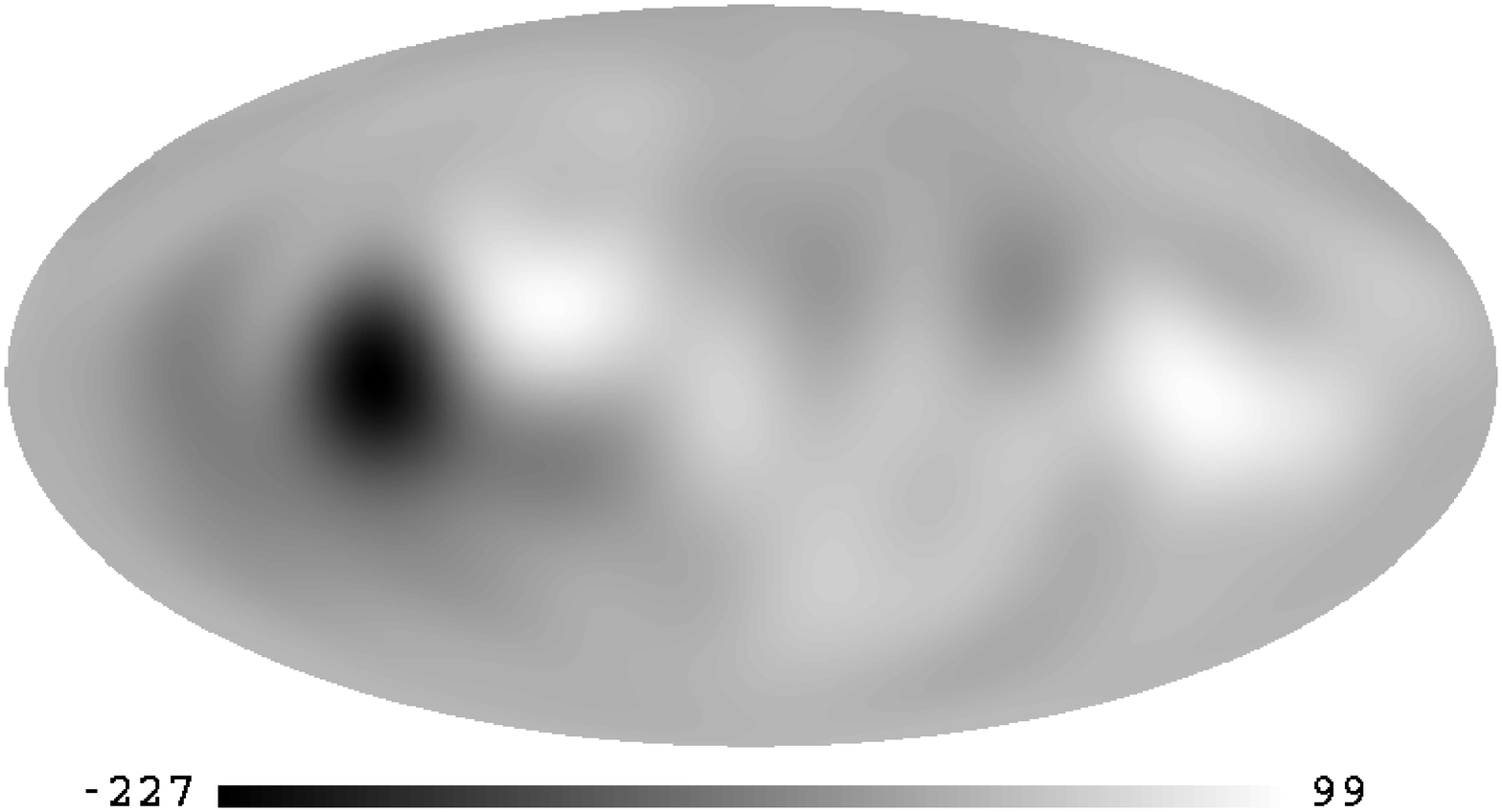,width=8cm}}\hfill
\centering{\epsfig{file=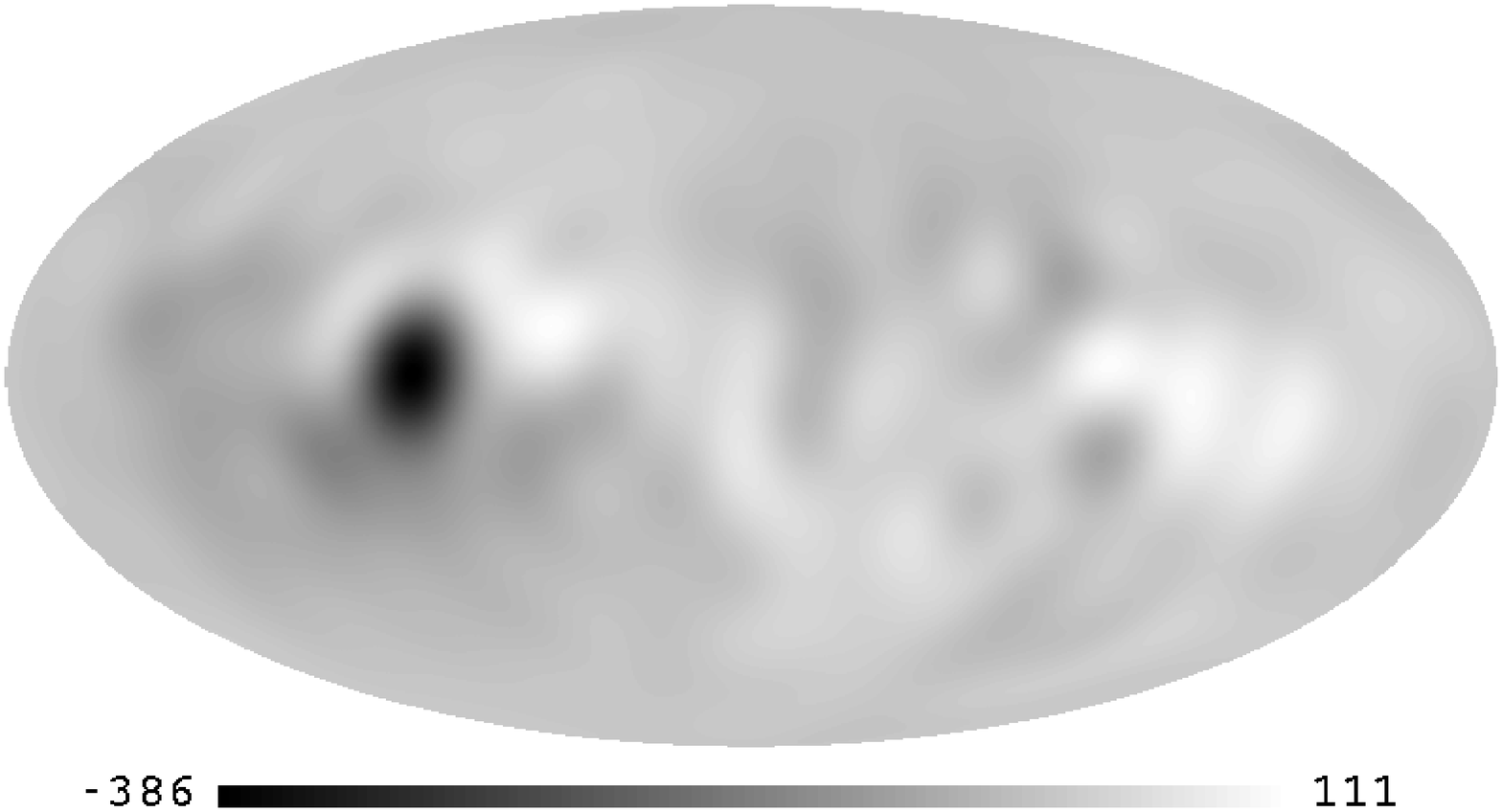,width=8cm}}
}
 \caption{\label{fig:simard} S81 catalogue with 540 sources. 
{\it Top}: $l_{\mathrm{max}}=$10. {\it Bottom}: $l_{\mathrm{max}}=$16. All maps are shown in Galactic coordinates with the Galactic centre in the middle and longitude increasing from right to left. The temperature-colour scales are measured in $\mathrm{rad\,m}^{-2}$.}
\end{figure}

\begin{figure} {
\centering{\epsfig{file=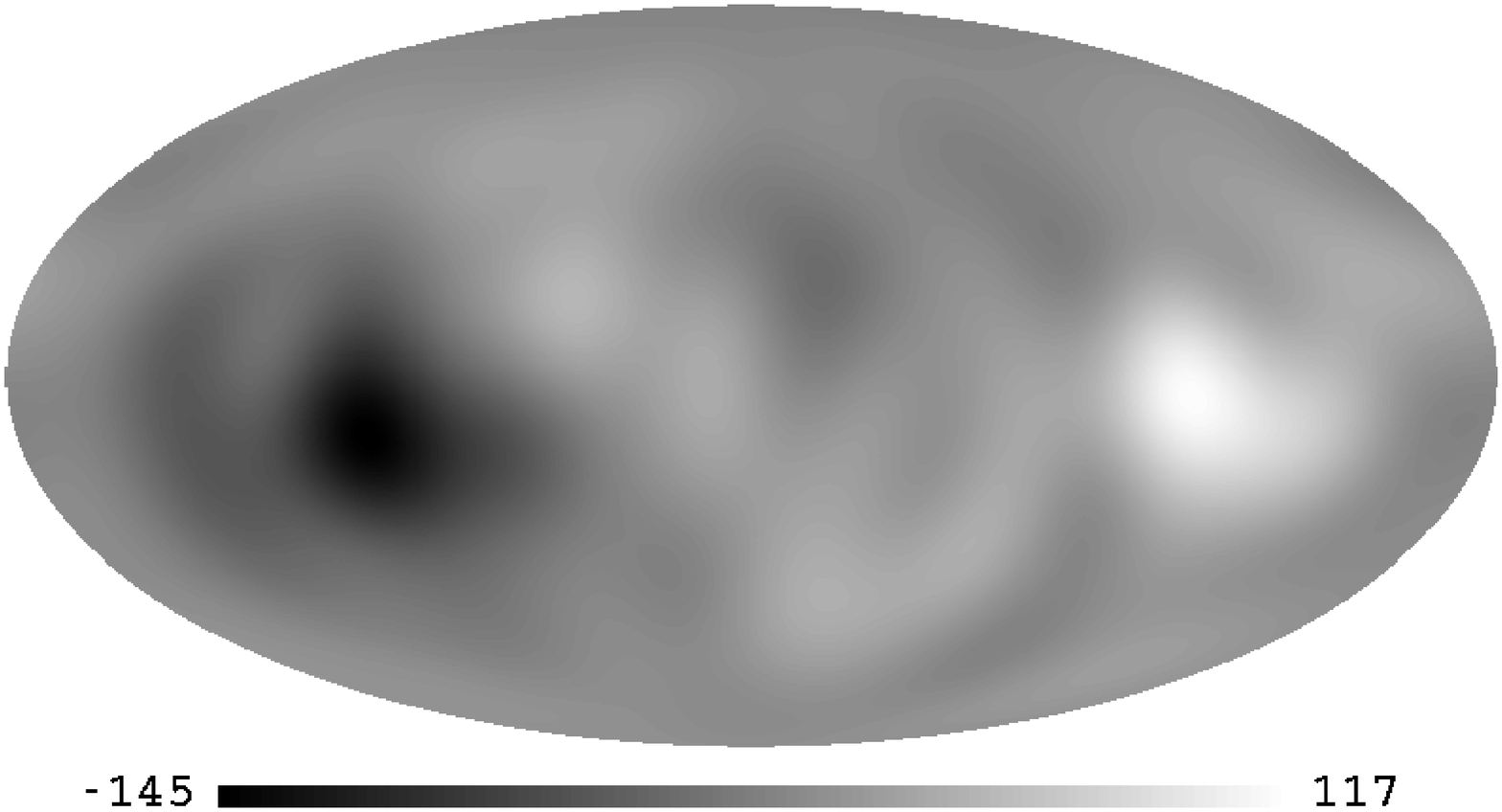,width=8cm}}\hfill
\centering{\epsfig{file=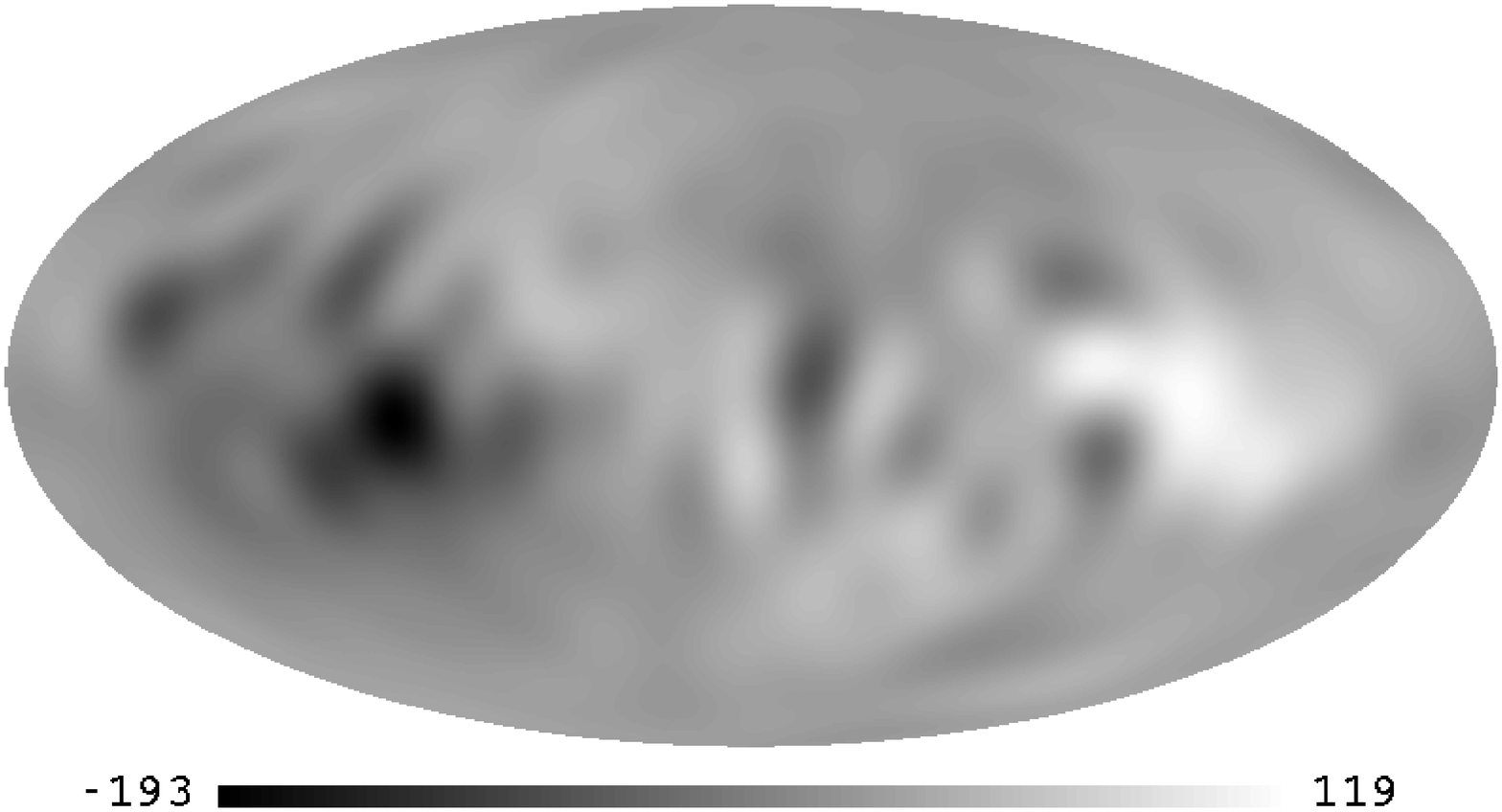,width=8cm}}
}
 \caption{\label{fig:broten} B88 catalogue with 644 sources. 
{\it Top}: $l_{\mathrm{max}}=$10. {\it Bottom}: $l_{\mathrm{max}}=$16.}
\end{figure}

\begin{figure} {
\centering{\epsfig{file=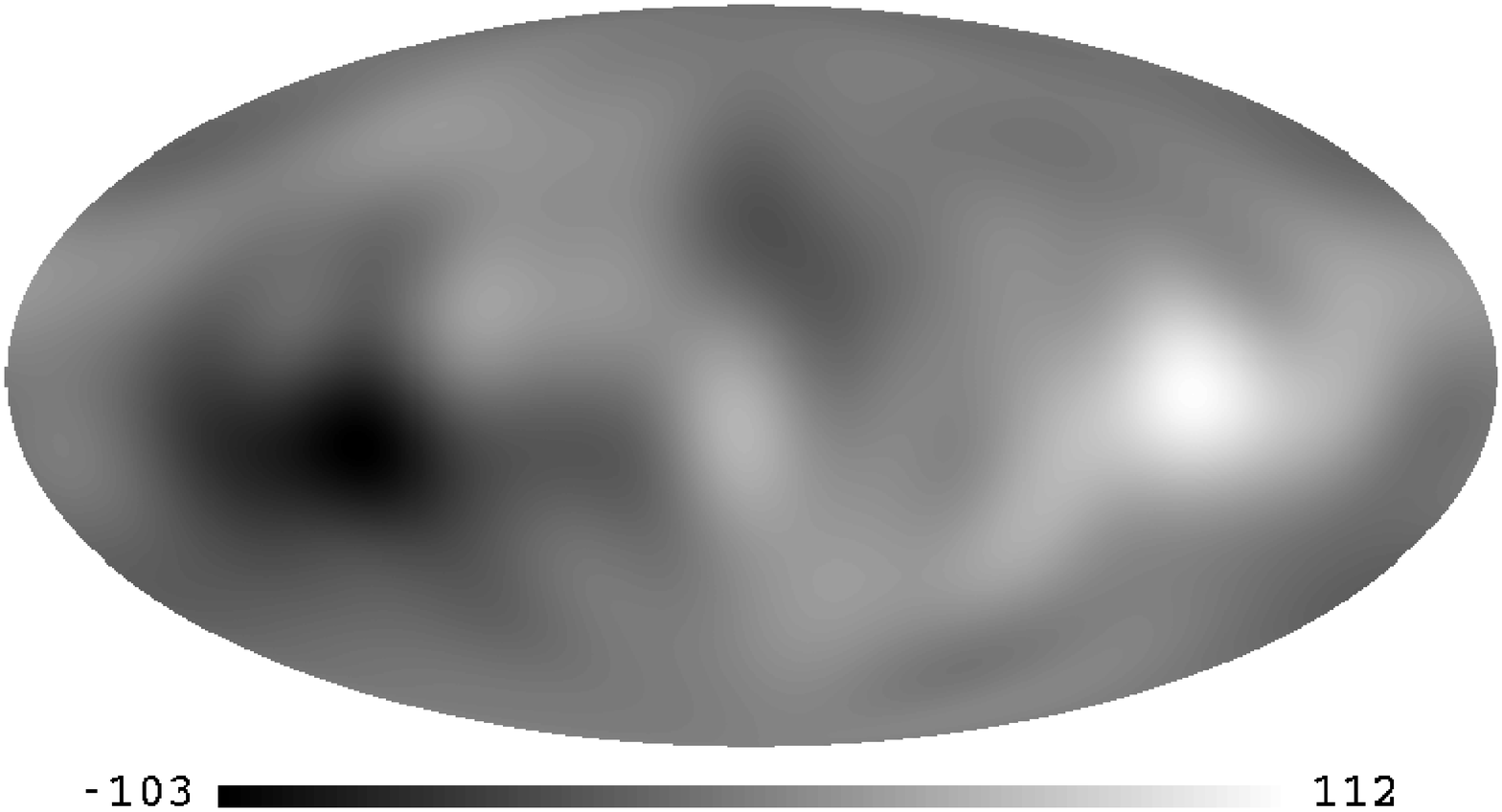,width=8cm}}\hfill
\centering{\epsfig{file=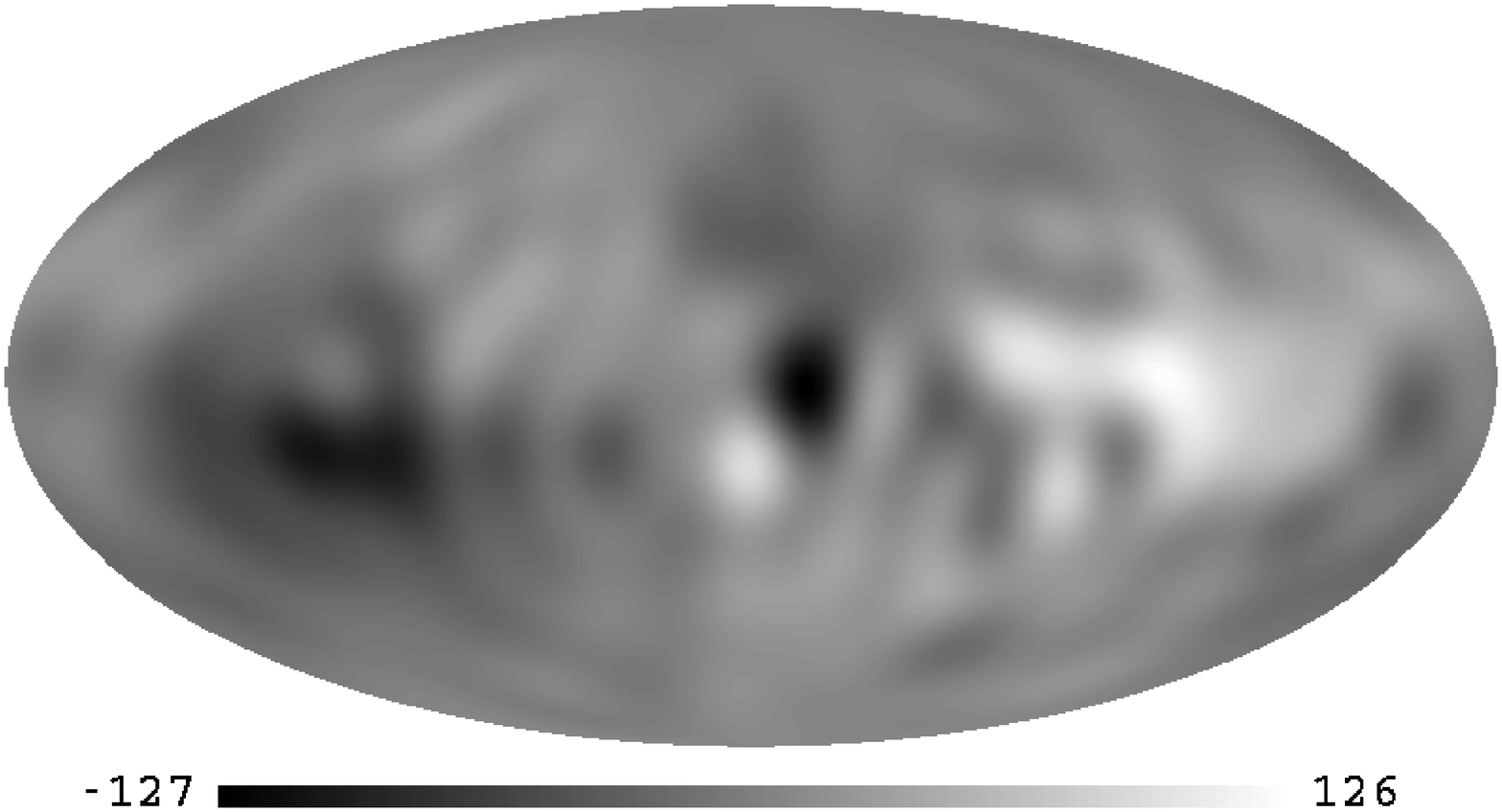,width=8cm}}
}
 \caption{\label{fig:rmjoint} F01 catalogue with 744 sources. 
{\it Top}: $l_{\mathrm{max}}=$10. {\it Bottom}: $l_{\mathrm{max}}=$16.}
\end{figure}

\begin{figure} {
\centering{\epsfig{file=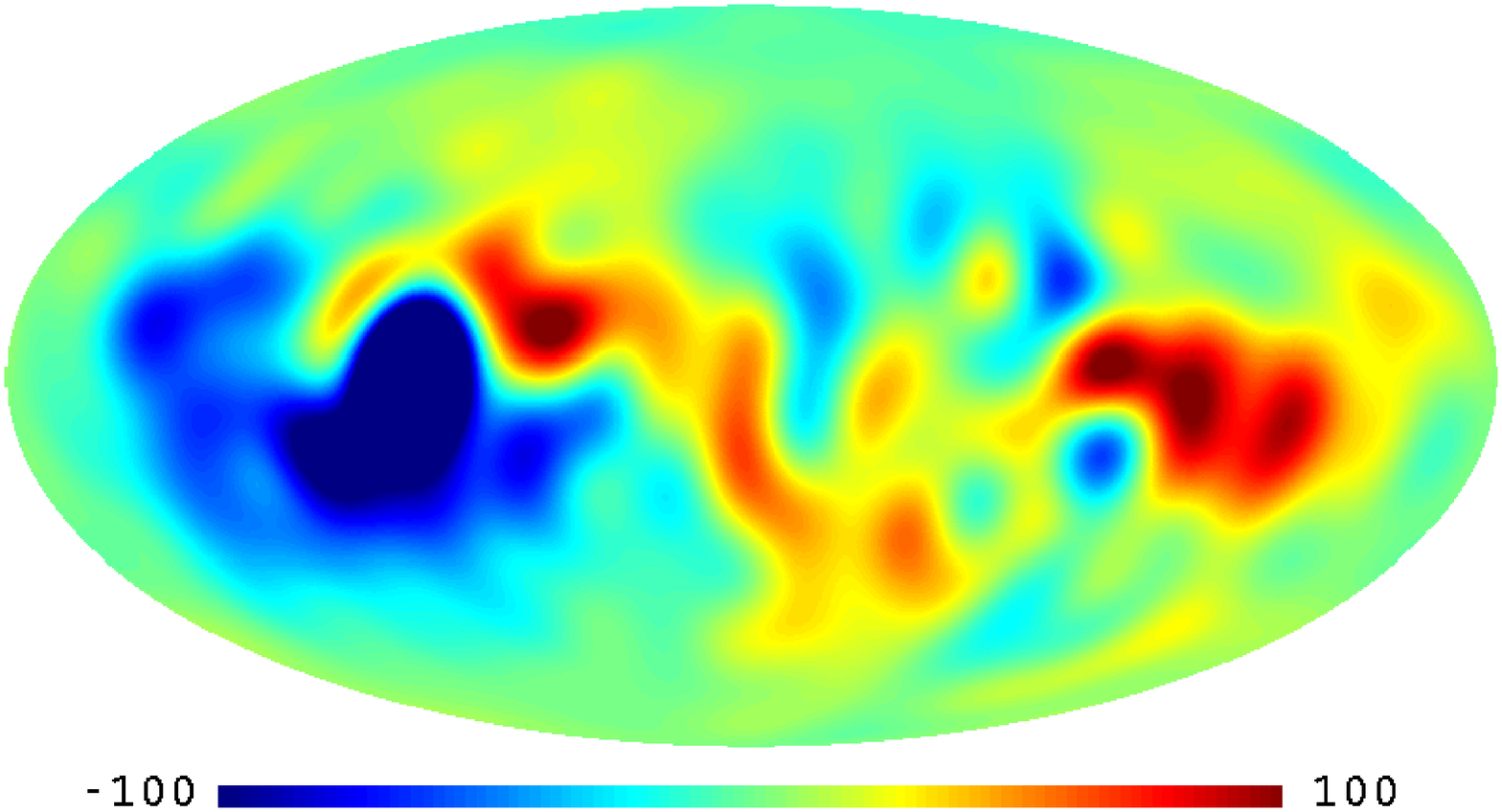,width=8cm}}\hfill 
\centering{\epsfig{file=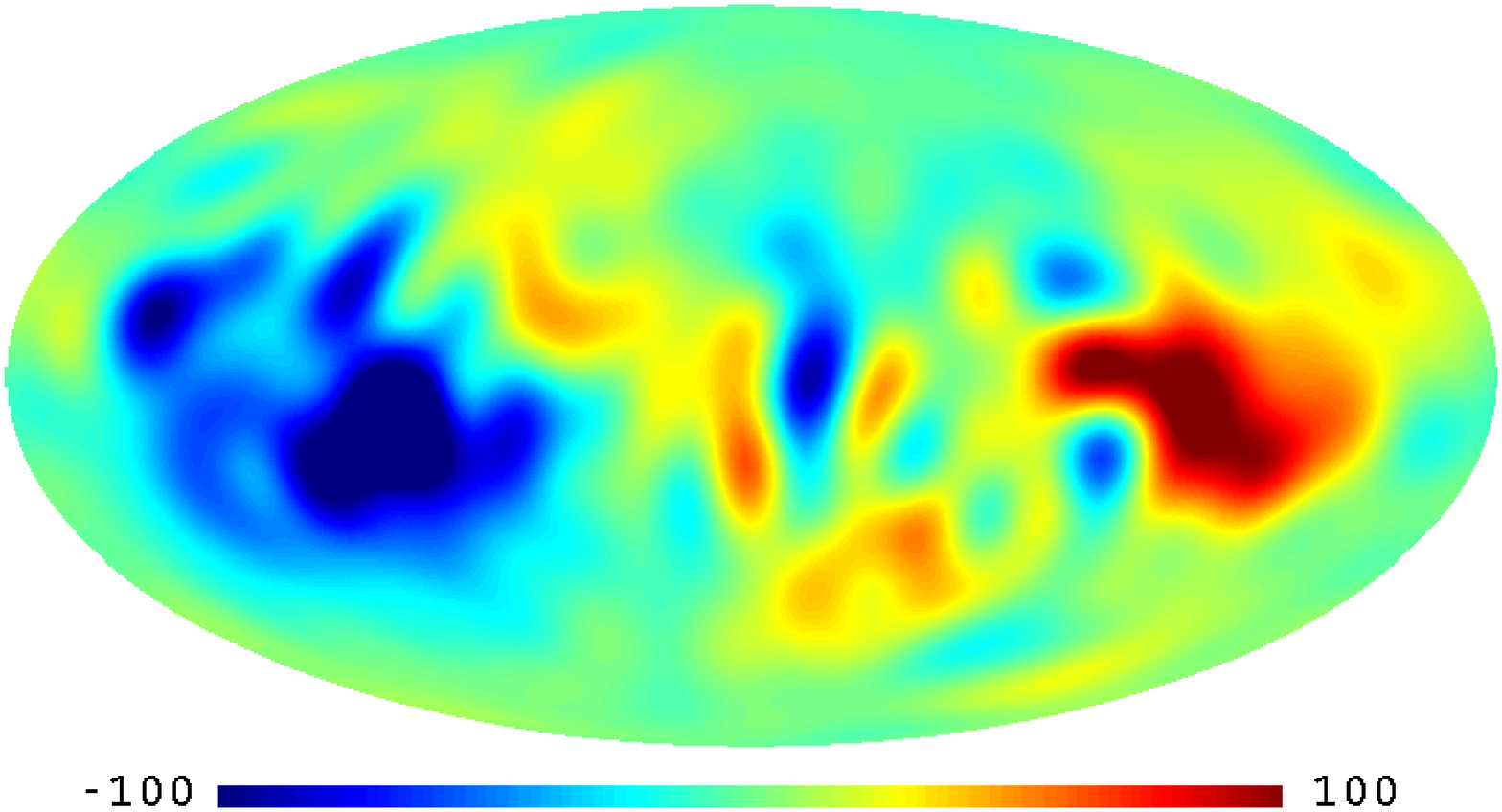,width=8cm}}\hfill
\centering{\epsfig{file=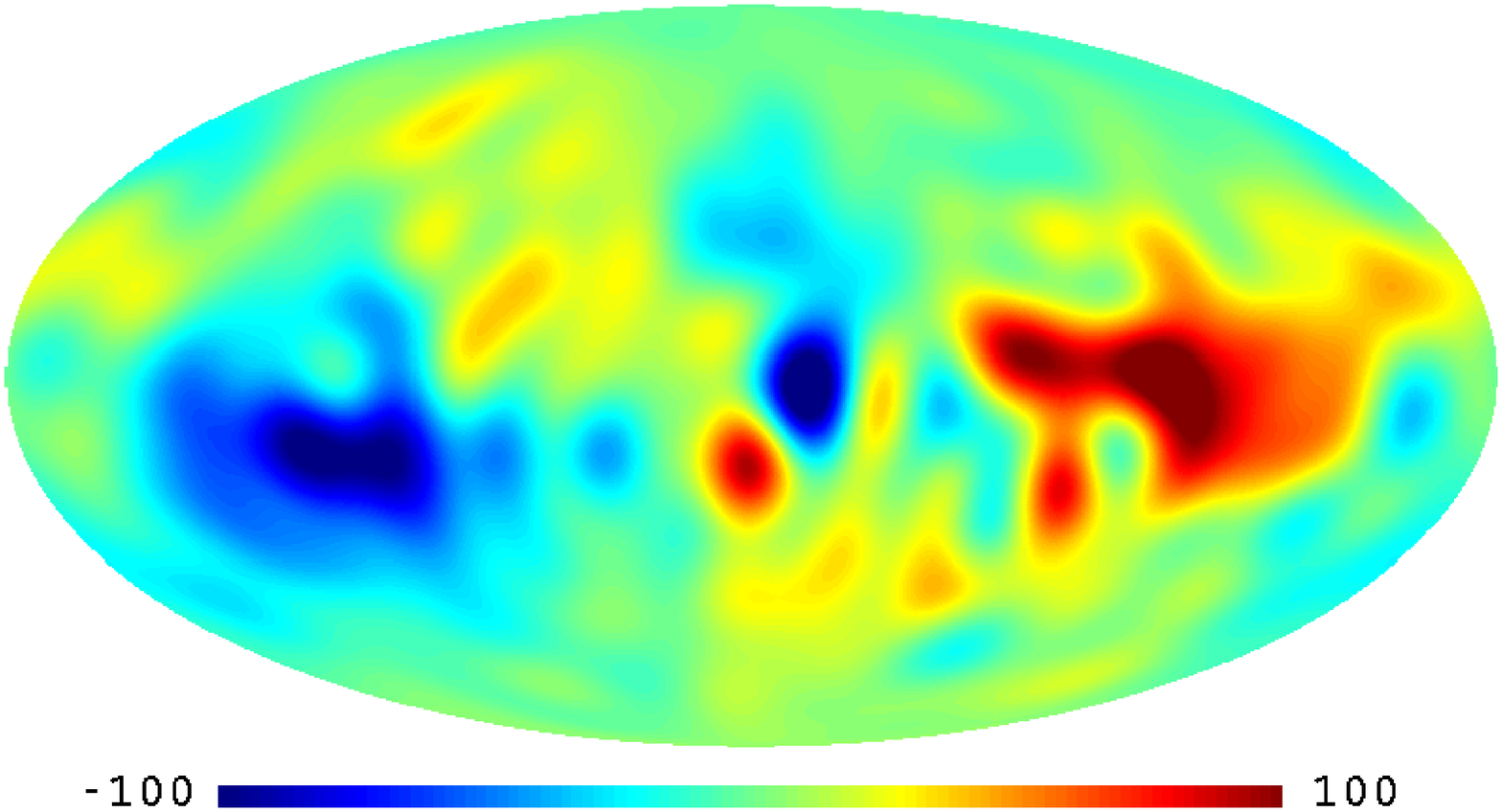,width=8cm}}
}
 \caption{\label{fig:uni} RM maps with identical
 temperature-colour scaling ($l_{\mathrm{max}}=16$). From {\it top} to {\it bottom}: S81, B88 and F01 catalogues.}
\end{figure}

In Figures \ref{fig:simard}, \ref{fig:broten} and \ref{fig:rmjoint}, we show 
the RM maps for the S81, B88 and F01 catalogues, respectively. We display only
the $l_{\mathrm{max}}$=10 and 16 maps in order not to overload the reader with
information. The r.m.s. values of $\cal R$ for the S81, B88 and F01 RM maps
($l_{\mathrm{max}}$=16) are
26.4, 23.5 and 21.5 $\mathrm{rad\,m}^{-2}$, respectively.  
Maxima (large positive ${\cal R}$) are white, whereas, minima (large
negative ${\cal R}$) are black. 
The two limiting scales enable us to see the progress of structure 
as the series expansion is extended to include higher modes. We can observe
how feature at small $l$ develop as the series extends. Reassuringly, the main features in
the $l_{\mathrm{max}}$=16 maps are also present in the $l_{\mathrm{max}}$=10 maps. 
The positions of the maxima and minima remain roughly unchanged. 
This is compelling evidence that the observed features are real. However,
comparison of the maps from the three catalogues is
inhibited by the temperature-colour scale varying from map to map. 
Therefore, for the $l_{\mathrm{max}}$=16 maps, we force the maximum and
minimum scale to be $|{\cal R}|=100\,\mathrm{rad\,m}^{-2}$ ; the results of which are shown in Figure \ref{fig:uni}. 

From Figure \ref{fig:uni}, it is clear that the maxima at $l^{\mathrm{II}}\!\sim\!270^{\mathrm{o}}$ and minima
at $l^{\mathrm{II}}\!\sim\!90^{\mathrm{o}}$ are
the dominant features in all three plots (here we use
$l^{\mathrm{II}}$ to denote Galactic longitude to avoid confusion with the
angular scale $l$). This maxima/minima pair corresponds to
the large-scale magnetic field in the local Orion spur (sometimes referred to
as an arm). These two spots are displaced from the equator to negative
Galactic coordinates. This asymmetry between the two hemispheres has been
widely reported before (eg. Vall\'ee \& Kronberg 1975; Frick et al. 2001); it
is usually attributed to the local radio Loop I (the North Galactic spur). Such
local distortions are associated with interstellar magnetised superbubbles
with typical diameters of 200 pc \cite{v97}.
There is also a prominent maxima/minima pair towards the Galactic centre in
the RM map
formed from the F01 catalogue. The centres of the maxima and minima are at
$l^{\mathrm{II}}=1^{\mathrm{o}}$ and $l^{\mathrm{II}}=346^{\mathrm{o}}$, respectively . On closer inspection, this
feature is present in the RM maps from the other two catalogues. Finally, in
the S81 RM map, there is a strong maxima at $l^{\mathrm{II}}\!\sim\!50^{\mathrm{o}}$ in the
northern hemisphere. This feature is only suggested in the other two maps.

\begin{figure} {
\centering{\epsfig{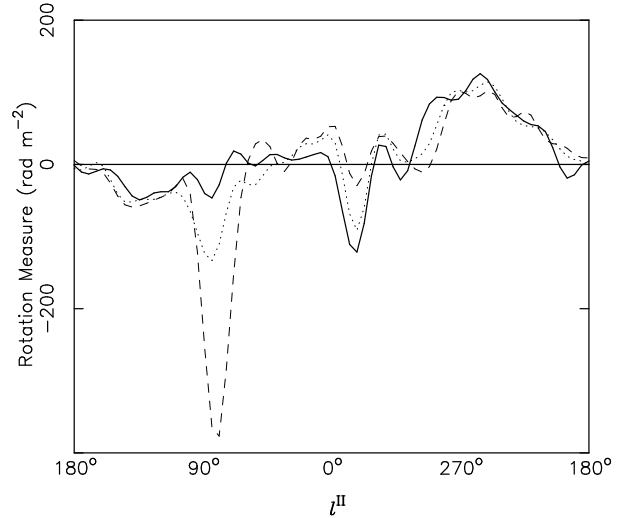}}
}
 \caption{\label{fig:equator} Galactic equator cross-section
 ($l_{\mathrm{max}}=16). ${\it Dashed}: S81 catalogue; {\it Dotted}: B88 catalogue; and {\it Solid line}: F01 catalogue}
\end{figure}

Cross-sections, along the Galactic equator, were taken of the RM maps
($l_{\mathrm{max}}$=16) in order to further understand the features. These are
shown in Figure \ref{fig:equator}. Ideally, with such a slice, maxima/minima
locations should indicate the tangential direction to spiral arms and directional field changes should correspond to
${\cal R}\!=\!0$. However, local distortions and flaws in the map-making
process, make this not entirely true. The orion spur location is clear for all
three maps. Furthermore, the maxima/minima pair towards the Galactic centre
(described in the previous paragraph) is
evident in all three cross-sections. However, the picture is hazy from $l^{\mathrm{II}}$=30-50$^{\mathrm{o}}$:
there is clear field reversal in S81 map; a hint of a reversal in F01 map;  and none in
B88 map. The maxima and minima in the cross-sections could be attributed to
the named inner arms (eg. the minima near the Galactic centre to the Norma arm), however, this
seems quite speculative given the variation from catalogue to catalogue. 

\begin{figure} {
\centering{\epsfig{file=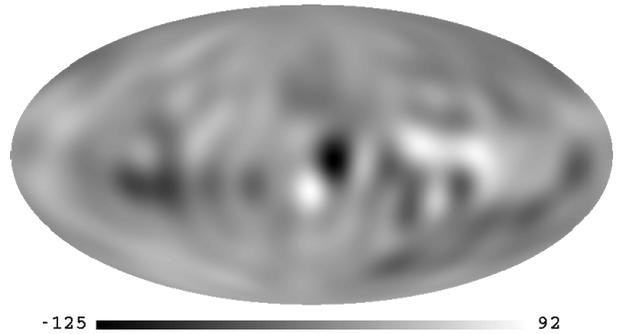,width=8cm}}
}
 \caption{\label{fig:nodipole} F01 catalogue with both the dipole and
 quadrupole removed ($l_{\mathrm{max}}=16$).}
\end{figure}

It is clear that the Orion spur is the dominant feature. Since the associated maxima/minima pair is separated by
180$^{\mathrm{o}}$, it will be the main source of the dipole ($l$=1). Moreover, we see
from the spectra that the quadrupole ($l$=2) is also strong. Therefore, we
remove both the dipole and quadruple from the RM map ($l_{\mathrm{max}}$=16) compiled from the F01
data. The results of which are displayed in Figure \ref{fig:nodipole}. This
enable us to view some of the smaller scale features. These details
will be hard to explain solely from Galactic magnetic field models. 
It will be interesting to see, if these small-scale features persist with
larger data sets.

A study of the global Galactic magnetic field structure would benefit
from the removal of local distortions. Consequently, we applied the method with the
region containing Loop I removed ($b\!>\!0^{\mathrm{o}}$, $0\!<\!l^{\mathrm{II}}\!<\!40^{\mathrm{o}}$,
$270\!<\!l^{\mathrm{II}}\!<\!360^{\mathrm{o}}$) \cite{rs77}. However, the removed segment was
too large to successfully reconstruct the sky given the remaining
sources. The lack of restrictions in the segment meant large maxima/minima
formed there. This highlights one disadvantage of spherical harmonic analysis over
wavelet analysis that can be localised in both physical and wavenumber
spaces. The removal of these local structures is useful for getting a clear picture of
the Galactic magnetic field. In CMB foreground studies, however, these
structures are essential components of a template.

We now turn to the question of errors in the derived RM maps. In
principle there are two distinct types of uncertainty that could
arise in the analysis we described above. The first concerns :
errors intrinsic to the measurement of ${\cal R}$ and the second
relates to errors resulting from sample selection.

We have tackled the first type of error by removing sources with
the most extreme values of ${\cal R}$ (those with ${\cal R} >
300\,\mathrm{rad\,m}^{-2}$), on the grounds that these are least
likely to be galactic. We expect the remaining experimental errors
to be stochastic and therefore the process of extracting
information on large--scales (as we do) should be unaffected by
any underlying noise. The spectrum of this noise should be flat
and only dominate on scales where the ``real'' power is weakest.
Experimental errors were addressed in some detail by Frick et al.
(2001) in the construction of their catalogue and their versions
of the two other catalogues used in our analysis. We feel it would
be inappropriate to repeat such a detailed analysis here.

\begin{figure} {
\centering{\epsfig{file=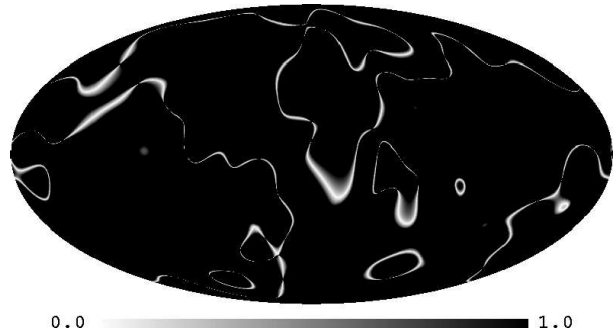,width=8cm}} }
\caption{\label{fig:sigtonoise} Signal--to--noise map constructed from 50 bootstrap samples of the F01 catalogue. Dark regions are
where the estimated signal--to--noise exceeds unity.}
\end{figure}

The second type of error corresponds to the sampling
uncertainties. This error could be computed reliably if we had a
large number of independent samples. This is, of course,
impossible but even in cases of non-repeatable observations there
are resampling techniques that can be used to make reasonable
estimates of these errors. The purpose of resampling the data is
to generate further sets with the same population distribution as
the original. Small perturbations to the original data set will
lead to this. For example, Ling, Frenk \& Barrow (1986) apply a
'bootstrap' resampling technique to estimate the sampling errors
in the two--point correlation function estimated from galaxy and
cluster redshift data. The bootstrap technique involves sampling
$N$ points (with replacement) from the original data set of $N$
sources in order to create pseudo data sets. The variation over an
ensemble of such resamplings is used to estimate the error in the
statistic in question for the original data. We applied this
method to the F01 catalogue with the intention of finding the
sample errors in the RM map with $l_{\mathrm{max}}$ set to 16. In
total, we produced 50 bootstrap samples from the original data set
and from each of these constructed a RM map. The resampling
technique is only designed to test the internal variance of the
true data set. Mean values obtained from the ensemble of pseudo
data sets are not expected to be good estimators of the true mean
values. Therefore, from the bootstrap samples, we calculate the
standard deviation $\sigma$ at each pixel position $p$
\begin{equation}\label{eqn:deviation}
\sigma(p)=\sqrt{ \sum_{i=1}^{50}\frac{{\cal R}_i(p)-\langle{\cal
R}(p)\rangle}{49}},
\end{equation}
where $i$ corresponds to the bootstrap sample. From this, we
constructed a signal--to--noise map where the signal is taken as
$|{\cal R}|$ and the noise is $\sigma$. This map is displayed in
Figure \ref{fig:sigtonoise}. The map saturates at a
signal--to--noise  of unity so the regions with low signal to
noise are more obvious. The mean signal--to--noise across the
whole sky is 117. We found this value to be very stable as the
number of bootstrap samples increased and this dictated the number
of pseudo data sets produced. Clearly, the majority of the sky is
unaffected by sample errors and we feel reassured that the
features seen are not the result of sample selection. It appears
that the boundaries between positive and negative regions of the
sky where the signal is correspondingly low are the most
susceptible to sample selection.

The spherical harmonic coefficients generated for all three
catalogues with $l_{\mathrm{max}}$ set to 16 are available at
\verb+http://www.nottingham.ac.uk/~ppxptd/rm_maps+. Hopefully,
this will enable our method to be compared with other techniques
and allow the maps to used in the investigation of observables
affected by the Galactic magnetic field. Instructions on the
generation of full--sky maps using the HEALPix package are also
given at this address.

\section{Correlations with CMB maps}
\label{sec:correlations}

As mentioned in the introduction, RM maps have a general importance
beyond trying to map the Galactic magnetic field. In what follows, we hope to
display one particular function. In this section, we focus solely on the RM map produced from
the F01 catalogue with $l_{\mathrm{max}}$ set to 16.
Dineen \& Coles (2004) developed a diagnostic of foreground contamination in
CMB maps. The method measured the cross-correlation between 
the RM of extragalactic sources and the observed
microwave signals at the same angular position. In what follows, we 
seek correlations between the spherical harmonic modes of the RM map and CMB-only
maps. In doing so, we shall look at the phases of the (complex) coefficients of
the modes from $l$=2 to $l$=16. Phase correlations have been
used before to hunt for evidence of departures in the CMB temperature field from
a Gaussian random field \cite{cdew04}. In Dineen,
Rocha \& Coles (2004) a certain form of phase correlation was found to be associated with non-trivial
topologies. Phase correlations between CMB and foreground maps
have been sought before \cite{ndv03,cn04}, however, here we wish to emphasise the virtue 
of having independent probes of Galactic foreground contamination. 

In order to seek evidence of phase correlations between the
RM map and CMB data, we turned to two WMAP-derived maps. Both 
were constructed in a manner that minimises foreground
contamination and detector noise, leaving a pure CMB signal. The
ultimate goal of these maps is to build an accurate image of the
last scattering surface (LSS) that captures the detailed morphology. Following the release
of the WMAP 1 yr data, the
 WMAP team \cite{bhhn03}, and Tegmark, de Oliveira-Costa \& Hamilton (2003; TOH) 
have released CMB-only sky maps (see papers for details).
We use the WMAP team's internal linear combination (ILC) map and the
Wiener-filtered map of TOH. The latter was chosen since the map was found to
be correlated with RM values in Dineen \& Coles (2004).

Two measures of phase association were used: the circular cross-correlation 
coefficient $R$ and Kuiper's statistic $V$. Both statistics will be evaluated
at each scale $l$ from 2 to 16. If we let $\Phi_{\mathrm{RM}}$ and 
$\Phi_{\mathrm{CMB}}$ be the phases of the RM and CMB maps, respectively. Then,
following Fisher (1993), $R$ is defined as:
\begin{equation}
R(l)=l^{-1}\sum_{m=1}^{l}\cos(\Phi_{m,\mathrm{RM}}-\Phi_{m,\mathrm{CMB}}).
\end{equation}
The expectation value of $R$ is 0, and hence highly correlated phases are
associated with large values of $|R|$. Kuiper's statistic 
is calculated from the available set of phase differences
($\Phi_{m,\mathrm{RM}}-\Phi_{m,\mathrm{CMB}}$) 
at a given scale. First, the phase differences are sorted into ascending order,
to give the set $\{\Theta _{1},\ldots ,\Theta _{p}\}$. Each angle
$\Theta _{j}$ is divided by $2\pi$ to give a set of variables
$X_{j}$, where $j=1\ldots p$. From the set of $X_j$ we derive two
values $S^+_{p}$ and $S^-_{p}$ where
\begin{equation}
S^{+}_{p} = {\rm max}
\left\{\frac{1}{p}-X_{1},\frac{2}{p}-X_{2},\ldots ,1-X_{p}\right\}
\end{equation}
and   \begin{equation} S^{-}_{p} = {\rm max}
\left\{X_{1},X_{2}-\frac{1}{p},\ldots
,X_{p}-\frac{p-1}{p}\right\}.
\end{equation}
Kuiper's statistic is then defined as
\begin{equation}
\label{TestStatisticV} V(l)=(S^{+}_{p}+S^{-}_{p})\cdot
\left(\sqrt{p}+0.155+\frac{0.24}{\sqrt{p}}\right).
\end{equation}
The form of $V$ is chosen so that it is approximately independent of 
sample size for large $p$.
Anomalously large values of $V$ indicate a distribution that is more 
clumped than a uniformly random distribution,
while low values mean that angles are more regular.

To access the significance of the values of $R$ and $V$ obtained from the
comparison of the RM map with the two CMB maps, we make use of Monte Carlo (MC) skies with
uniformly random phases. The statistics were calculated for 10,000 MC
skies contrasted with a further 10,000 MC skies. Thus, we are left with 10,000 values of $R$ and
$V$ for each scale. 

\begin{figure} {
\centering{\epsfig{file=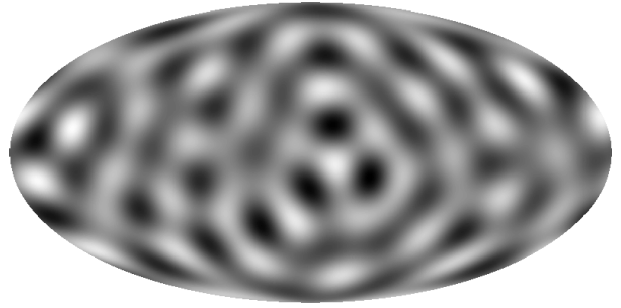,width=8cm}}
}
 \caption{\label{fig:ilc11} Internal linear combination map constructed with
 $a_{l,m}$ for $l$=11 only.}
\end{figure}

The results from both CMB maps suggests that there is strong correlations between
the phases at $l$=11. For the ILC, the values of $R(11)$ and $V(11)$ are greater than 99
percent of the MC values. Whereas, the values of $R(11)$ and $V(11)$ corresponding to
the TOH Wiener-filtered map are greater than 97 and 98 percent of the MC
skies, respectively. In Figure \ref{fig:ilc11} we plot the ILC map constructed
with the $a_{l,m}$ for $l$=11. Overlapping this image with that of the RM in
Fig \ref{fig:rmjoint}, we can see that the central maxima/minima pair
in the RM map are similar in location, size and shape to structure in the ILC image
(but with colours reversed). This is probably what determines the specific
scale $l$=11. 
Interestingly, $l$=11 corresponds to the scale 
that Naselsky, Doroshkevich \& Verkhodanov (2003) found the greatest
level of correlation between the ILC phases and those of the foreground maps.

\section{Conclusion}
\label{sec:conclusion}

In this paper, we have presented a new method to generate all-sky RM maps from uneven
and sparsely populated data samples. The method calculates a set of
functions orthonormal to the data set. With these basis functions, the
spherical harmonic coefficients are calculated and converted into sky maps using the HEALPix
package. The method was applied to three catalogues; S81, B88 and F01 catalogues. Maps from each catalogue
showed evidence of the magnetic field in our local Orion spur, the North-South
asymmetry attributed to radio Loop I and a maxima/minima pair close to the Galactic
centre that possibly corresponds to
the magnetic field of two inner Galactic arms. A RM map constructed from the S81 catalogue
also had a prominent maxima at $l^{\mathrm{II}}\!\sim\!50^{\mathrm{o}}$ in the northern hemisphere.

In Section \ref{sec:correlations}, we showed the benefits a RM map has to CMB foreground analysis.
Phase correlations were sought between RM maps and those of CMB-only maps
derived from the WMAP data. For both the WMAP team's ILC map and the
Wiener-filtered map of TOH, phases corresponding to $l$=11 were found to be
highly correlated. Naselsky, Doroshkevich \& Verkhodanov (2003) found the same
scale to display phase correlations when carrying out a similar analysis on
the ILC map and foreground maps. Their detection of correlations at the same scale as our analysis, reaffirms that the
RM catalogues are valuable independent tracers of CMB foregrounds \cite{dc04}.

Modelling foregrounds will play a crucial role in CMB polarisation
studies. Foreground contamination is expected to be more severe than in the temperature
measurements \cite{k99}. Consequently, superior templates for the individual foreground
components are required. RM maps will help trace these components. 
Besides this, extrapolation of low frequency measurement of synchrotron
polarisation to CMB frequencies has been shown to be complicated by Faraday rotation
\cite{dtok03}. Again, this underlines the importance of 
developing templates of the Faraday rotation of the Galactic sky.
Efforts to map the RM sky will be greatly enhanced by increased source
catalogues for both extragalactic sources and pulsars within our Galaxy. This
may enable the formation of a 3-dimensional image of the Galactic magnetic
field. Furthermore, attempts to map the RM sky will be enhanced by upcoming
satellite CMB polarisation experiments which present unprecedented sky-coverage and resolution.

\section*{Acknowledgements}
We thank Anvar Shukurov and Rodion Stepanov for providing us with the rotation
measure catalogues and useful comments. We
gratefully acknowledge the use of the HEALPix package and the Legacy
Archive for Microwave Background Data Analysis (LAMBDA). Support
for LAMBDA is provided by the NASA Office of Space Science.


\begin{thebibliography}{}
\bibitem[Barrow, Ferreira \& Silk 1997]{bfs97} Barrow J.D., Ferreira P.G. \&
  Silk J., 1997, Phys. Rev. Lett., 78, 3610
\bibitem[Bennett et al. 1994]{bkhb94} Bennett C.L. et al., 1994, ApJ, 436, 423
\bibitem[Bennett et al. 2003]{bhhn03} Bennett C.L. et al., 2003, ApJS, 148, 97
\bibitem[Broten et al. 1988]{bmv88} Broten N.W., MacLeod J.M. \& Vall\'ee
  J.P., 1988, Ap\&SS, 141, 303
\bibitem[Chen et al. 2004]{cmkr04} Chen G., Mukherjee P., Kahniashvili T., Ratra B. \&
 Wang Y., 2004, ApJ, 611, 655
\bibitem[Chiang \& Naselsky 2004]{cn04} Chiang L.-Y. \& Naselsky P., 2004, astro-ph/0407395
\bibitem[Coles et al. 2004]{cdew04} Coles P., Dineen P., Earl J. \& Wright D.,
  2004, MNRAS, 350, 983 
\bibitem[de Oliveira-Costa et al. 2003]{dtok03}de Oliveira-Costa A., Tegmark M.,
  O'Dell C., Keating B., Timbie P., Efstathiou G. \&  Smoot G., 2003,
  Phys. Rev. D, 68, 83003
\bibitem[de Oliveira-Costa et al. 2004]{dtdg04} de Oliveira-Costa A., Tegmark M., Davies R.D.,
  Guti\'errez C.M., Lasenby A.N., Rebolo R. \& Watson R.A., 2004, ApJ, 606, L89
\bibitem[Dineen \& Coles 2004]{dc04} Dineen P. \& Coles P., 2004, MNRAS, 347,
  52
\bibitem[Dineen, Rocha \& Coles 2004]{drc04} Dineen P., Rocha G. \& Coles P.,
  2004, submitted to MNRAS, astro-ph/0404356
\bibitem[Draine \& Lazarian 1998]{dl98} Draine B.T. \& Lazarian A., 1998, ApJ,
  494, L19
\bibitem[Fisher 1993]{fish} Fisher N.I., 1993, Statistical Analysis of Circular Data. Cambridge University Press, Cambridge.
\bibitem[Frick et al. 2001]{fsss01} Frick P., Stepanov R., Shukurov A. \&
  Sokoloff D., 2001, MNRAS, 325, 649
\bibitem[G\'orski 1994]{g94} G\'orski K.M., 1994, ApJ, 430, L85
\bibitem[G\'orski, Hivon \& Wandelt 1998]{healpix} G\'orski K.M.,
Hivon E. \&  Wandelt B.D., 1999, in Proceedings of the MPA/ESO
Conference {\em Evolution of Large-Scale Structure}, eds. A.J.
Banday, R.S. Sheth and L. Da Costa, PrintPartners Ipskamp, NL, pp.
37-42 (also astro-ph/9812350)
\bibitem[Han 2004]{h04} Han J.L., 2004, astro-ph/0402170
\bibitem[Kim, Tribble \& Kronberg 1991]{ktk91} Kim K.-T., Tribble P.C., 
Kronberg P.P. 1991, ApJ, 379, 80
\bibitem[Kogut et al. 2003]{ksbb03} Kogut A. et al., 2003, ApJS, 148, 161
\bibitem[Kosowsky 1999]{k99} Kosowsky A., 1999, New Astronomy Reviews, 43, 157
\bibitem[Kosowsky \& Loeb 1996]{kl96} Kosowsky A. \& Loeb A., 1996, ApJ, 469, 1 
\bibitem[Kovac et al. 2002]{klpc02} Kovac J. M., Leitch E. M., Pryke C., Carlstrom J. E., Halverson N. 
W. \&  Holzapfel W. L., 2002, Nature, 420, 772
\bibitem[Kronberg, Perry \& Zukowski 1992]{kpz92} Kronberg P.P., Perry J.J. \&
  Zukowski E.L., 1992, ApJ, 387, 528
\bibitem[Lewis 2004]{l04} Lewis A., 2004, accepted by Phys. Rev. D, astro-ph/0406096
\bibitem[Ling, Frenk \& Barrow 1986]{lfb86} Ling E.N., Frenk C.S. \& Barrow J.D., 1986, MNRAS, 223, 21
\bibitem[Naselsky, Doroshkevich \& Verkhodanov 2003]{ndv03} Naselsky P.,
  Doroshkevich A. \& Verkhodanov O., 2003, ApJ, 599, L53 
\bibitem[Naselsky et al. 2004]{ncov04} Naselsky P.D., Chiang L.-Y., Olesen
  P. \&  Verkhodanov O.V., 2004, accepted by ApJ, astro-ph/0405181
\bibitem[Ruzmaikin \& Sokoloff 1977]{rs77} Ruzmaikin \& Sokoloff, 1977, A\&A,
  58, 247
\bibitem[Scannapieco \& Ferreira 1997]{sf97} Scannapieco E.S. \& Ferreira
  P.G., 1997, Phys. Rev. D, 56, R7493
\bibitem[Sc\'occola, Harari \& Mollerach 2004]{shm04} Sc\'occola C., Harari D. \&
  Mollerach S., 2004, astro-ph/0405396
\bibitem[Seymour 1966]{s66} Seymour P.A.H., 1966, MNRAS, 134, 389
\bibitem[Seymour 1984]{s84} Seymour P.A.H., 1984, QJRAS, 25, 293
\bibitem[Simard-Normandin et al. 1981]{skb81} Simard-Normandin M., Kronberg
  P. \& Button S., 1981, ApJS, 45, 97
\bibitem[Sofue \& Fujimoto 1983]{sf83} Sofue Y. \& Fujimoto M., 1983, ApJ,
  265, 722
\bibitem[Tegmark, de Oliveira-Costa \& Hamilton 2003]{tdh03} Tegmark M., 
de Oliveira-Costa A. \& Hamilton A., 2003, Phys. Rev. D, 68, 123523
\bibitem[Vall\'ee 1997]{v97} Vall\'ee J.P., 1997, Fundamentals of Cosmic
  Physics, 19, 1
\bibitem[Vall\'ee \& Kronberg 1975]{vk75} Vall\'ee J.P. \& Kronberg P.P., 
1975, A\&A, 43, 233
\bibitem[Wasserman 1978]{w78} Wasserman I., 1978, ApJ, 224, 337
\bibitem[Widrow 2002]{w02} Widrow L.M., 2002, Rev. Mod. Phys., 74, 775
\bibitem[Wolfe, Lanzetta \& Oren 1992]{wlo92} Wolfe A.M., Lanzetta K.M, \&
  Oren A.L., 1992, ApJ, 388, 17
\end{thebibliography}
\end{document}